\documentclass[11pt]{book}
\usepackage{a4wide}
\usepackage{fancyheadings}
\usepackage{setspace}
\usepackage{newcent}
\usepackage{xspace}
\usepackage{amssymb}
\usepackage{amstext}

\newcommand{\degr}  {^{\circ}}

\newcommand{\farcs} {\rlap.{''}}
\newcommand{\fdg}   {\rlap.^{\circ}}

\begin{document}
\thispagestyle{empty}

\begin{center}
  \subsection*{University of Barcelona}

  \subsubsection*{PhD Thesis in Astronomy}

  \section*{New observational techniques and analysis tools for wide field CCD surveys and high resolution astrometry}

  \subsubsection*{by}
  \section*{Octavi Fors Aldrich}
  \subsubsection*{March 7th, 2006}
\end{center}

\vspace{1cm}

\section*{Abstract}

\noindent The aim of this thesis is two-fold. First it provides a general
methodology for applying image deconvolution to wide-field CCD imagery. Second,
two new CCD observational techniques and two data analysis tools are proposed
for the first time in the context of high resolution astrometry, in particular
for lunar occultations and speckle interferometry observations.\\

In the first part of the thesis a wavelet-based adaptive image deconvolution
algorithm ({\tt AWMLE}) has been applied to two sets of survey type CCD data:
QUasar Equatorial Survey Team project ({\tt QUEST}) and Near-Earth Space
Surveillance Terrestrial ({\tt NESS-T}). Richardson-Lucy image deconvolution has
also been used with Flagstaff Transit Telescope ({\tt FASTT}) imagery. Both the
obtaining and performance of those images were accomplished by following a new
methodology which includes accurate image calibration, source detection and
centering, and correct assessment procedures of the performance of the
deconvolution. Results show that {\tt AWMLE} deconvolution can increase limiting
magnitude up to 0.6 mag and improve limiting resolution 1 pixel with respect to
original image. These studies have been conducted in the context of programs
dedicated to macrolensing search ({\tt QUEST}) and NEOs discovery ({\tt
NESS-T}). Finally, astrometric accuracy of {\tt FASTT} images have not been
found to change significantly after deconvolution. In the same way, no
positional bias towards the centre of the pixel has been observed.\\

In the second part of the thesis a new observational technique based on CCD fast
drift scanning has been proposed, implemented and assessed for lunar
occultations (LO) and speckle interferometry observations. 

In the case of LO, the technique yielded positive detection of binaries up to 2~mas of projected separation and stellar diameters measurements in
the 7~mas regime. The proposed technique implies no optical or
mechanical additional adjustments and can be applied to nearly all available
full frame CCDs. Thus, it enables all kind of professional and high-end amateur
observatories for LO work. 

Complementary to this work, a four-year LO program (CALOP) at Calar Alto
Observatory spanning 71.5 nights of observation and 388 recorded events has been
conducted by means of CCD and MAGIC IR array cameras at OAN 1.5m and CAHA 2.2m
telescopes. CALOP results include the detection of one triple system and 14 new
and 1 known binaries in the near-IR, and one binary in the visible. Their
projected separations range from 90 to 2~mas with brightness ratios up to 1:35
in the $K$ band. Several angular diameters have been also measured in the
near-IR. The performance of CALOP has been calibrated in terms of limiting
magnitude ($K_{lim}\sim9.0$) and limiting angular resolution ($\phi_{\rm
lim}\sim$1-3~mas). In addition, the binary detection probability of the program
has found to be about 4\%. Finally, a new wavelet-based method for extracting
and characterizing LO lightcurves in an automated fashion was proposed,
implemented and applied to CALOP database. This pipeline addresses the need of
disposing of preliminary results in immediate basis for future programs which
will provide larger number of events.

In the case of speckle interferometry, CCD fast drift scanning technique has
been validated with the observation of four binary systems with well determined
orbits. The results of separation, position angle and magnitude difference are
in accordance with published measurements by other observers and predicted
orbits. Error estimates for these have been found to be $0\farcs017$, $1\fdg5$ and 0.34 mag, respectively. These are in the order of other authors and
can be considered as successful for a first trial of this technique.
Finally, a new approach for calibrating speckle transfer function from the
binary power spectrum itself has been introduced. It does not require point
source observations, which gives a more effective use of observation time. This
new calibration method appears to be limited to zenith angles above
$30\degr$ when observing with no refraction compensation devices.

\end{document}